\documentclass[12pt,draftcls,a4paper,onecolumn]{IEEEtran}
\ifCLASSINFOpdf
  
\else
 
\fi

\ifCLASSOPTIONcompsoc
    \usepackage[caption=false, font=normalsize, labelfont=sf, textfont=sf]{subfig}
\else
    \usepackage[caption=false, font=normalsize]{subfig}
\fi
\usepackage{lipsum}%

\usepackage{balance}
\usepackage{multicol}   
\usepackage{cite}
\usepackage{gensymb}
\usepackage{multirow}
\usepackage{graphics}  
\usepackage{epsfig} 
\usepackage{graphicx}
\usepackage{epstopdf}
\usepackage{textcomp}
\usepackage{amsmath}
\usepackage{mathtools}
\usepackage{filecontents}
\usepackage{lipsum,color}
\usepackage{mathtools,tikz,caption}
\usepackage{amssymb}
\usepackage{amsfonts}
\usepackage{float,stfloats,blindtext}
\definecolor{teal}{rgb}{0.0, 0.0, 0.0}
\usepackage[T1]{fontenc}
\begin{document}
	
\title{\Large Beam Tracking for UAV-Assisted FSO Links With a Four-Quadrant Detector}

\author{Hossein~Safi,~Akbar~Dargahi,~and~Julian~Cheng,~\IEEEmembership{Senior Member, IEEE}


\thanks{H. Safi and A. Dargahi are with the Department of Electrical Engineering, Shahid Beheshti University G. C., 1983963113, Tehran, Iran (e-mail: \{h$\_$safi, a$\-$dargahi\}@sbu.ac.ir). J .Cheng is with the School of Engineering, The University of British Columbia, V1V 1V7, Kelowna, BC, Canada (email:julian.cheng@ubc.ac)}}


\maketitle
\begin{abstract}
A ground-to-air free-space optical link is studied for a hovering unmanned aerial vehicle (UAV) having multiple rotors. \textcolor{teal}{For this UAV, a four-quadrant array of photodetectors  is used at the optical receiver to alleviate the adverse effect of  hovering fluctuations by enlarging the receiver field-of-view.} Extensive mathematical analysis is  conducted to evaluate the beam tracking performance under the random effects of hovering fluctuations. \textcolor{teal}{The
	accuracy of the derived analytical expressions is corroborated by
	performing Monte-Carlo simulations.  It is
	shown that the performance of such links depends heavily on the random fluctuations of hovering UAV, and, for each level of instability there is an optimal size for the array that minimizes the tracking error probability}
\end{abstract}


\IEEEpeerreviewmaketitle
\section{Introduction}
\label{I}
Recently, unmanned aerial vehicle (UAV) communications have attracted significant interest due to
many advantageous such as fast deployment, flexible configuration, and possibility of having better channel conditions. 
 Furthermore, technological advances of UAVs along with their reduced production costs have made future cellular networks more likely to be equipped with UAVs as flying base stations (BSs) \cite{8796414}. Nevertheless, existing terrestrial wireless networks may experience radio interference when UAVs are used as aerial BSs, making it challenging to implement a practical aerial RF link. UAVs and free-space optical (FSO) based communicaiton systems have been recently proposed as a promising approach for the next generation  wireless network \cite{alzenad2018fso}, where the aerial communication links are free of radio interference.  FSO links offer higher bandwidth and security than the conventional RF links, and thus, they can play
a pivotal role in supporting emerging data-hungry applications. It is of great importance for the optical  transmitter in a UAV-assisted FSO link to point precisely at the receiver field-of-view (FoV) in order to avoid  beam misalignment. However, when optical subsystems are mounted on a hovering UAV, orientation deviations (due to the UAV\textquotesingle s random wobbling) can cause angle of arrival (AoA) fluctuations of optical beam at the lens aperture, and, subsequently, image beam will dance around the area of photodetector (PD) \cite{safi2020analytical}. Hence, for an optical link from a ground node towards a UAV-based receiver,  it becomes unavoidable to perform beam tracking to determine the direction of arrival beam at the receiver plane and re-establish  the link in the case of beam misalignment. 

For terrestrial FSO links where transceivers are firmly fixed, the effect of AoA fluctuations can be neglected. Indeed, most of the literature in the context of optical beam tracking has been dedicated to space optics, i.e., inter-satellite links and earth-space laser communication \cite{gagliardi1995optical}. Moreover, an extensive survey was conducted on different acquisition, tracking, and pointing (ATP) methods for mobile FSO communications, and different use cases of those methods were categorized \cite{kaymak2018survey}. However, most of the existing methods require heavy and bulky mechanical or piezoelectric
equipment, e.g., gimbals and retro reflectors, which are  inappropriate for small-sized multi-rotor UAVs with limited payload.

 \textcolor{black}{To relax practical constraints imposed by payload and power consumption limitations while combating the adverse effects of AoA fluctuations,  a practical and cost-effective solution is to replace the ATP unit on the small UAV with an array of PDs \cite{gagliardi1995optical, zhang2019improved}, which gives three practical advantageous. First, the receiver FoV can be made wider to
 	compensate the adverse effect of AoA fluctuations on the system performance. Second, the sophisticated ATP subsystem is moved to the optical transceiver on the ground station. Third, the instantaneous orientation errors can be fixed by estimating the position of the beam on the array (i.e., the AoA of the beam), and feeding it back through a control message
 	to the mechanical subsystem of the UAV\footnote{\textcolor{black}{Here, we assume that the UAV orients itself or use a simple servo motor (which
 		has much lower weight and price than a bulky stabilizer) to align detector plane towards the ground transmitter.}}.} 
 	
 	\textcolor{black}{Optical beam tracking using detector arrays has recently been studied assuming  the photon-counting regime, e.g., a deep-space optical
 	communication setting \cite{bashir2017optical, bashir2019free, bashir2021cramer,bashir2020beam}. For instance, the photodetection was modeled by a non-homogeneous Poisson process \cite{bashir2017optical}, and two filtering methods, i.e., a Kalman filter and  a particle filter, were proposed for tracking the beam position. In addition, a mathematical analysis was performed
 	to show that the probability of error decreases monotonically as the number of PD
 	in the array is increased \cite{bashir2019free}. Meanwhile, the beam position estimation
 	problem was examined \cite{bashir2021cramer} for photon-counting detector arrays and the Cramér-Rao lower bounds were derived for the variance
 	of unbiased estimators. Furthermore, beam tracking algorithms were proposed \cite{bashir2020beam} using received photon counts during an observation interval as
 	a sufficient statistic for tracking the beam position.}

\textcolor{teal}{Nevertheless, the results of these prior works are applicable to photon-limited channels of deep space
 or long-range FSO communications in which the long link distances reduce the number of received signal photons significantly. Although such systems can deliver a better performance than
 conventional ones with direct detection, but at the expense of higher implementation cost.}
\textcolor{teal}{In addition, if a UAV-based FSO system is considered as an aerial access point in the
	backhaul or fronthaul link of next generation wireless networks, the received optical power must
	be high enough to accommodate the high quality-of-service demand of such systems, e.g., bit-error
	rate (BER) lower than $10^{-9}$. In such a high received power regime, the output of the
	photodetectors is well approximated by a Gaussian distribution \cite{zabih}.}.
 
 \textcolor{teal}{The main contribution of this letter is the mathematical analysis carried out to derive
 	the tracking error probability when a four-quadrant array of PDs is used at the optical receiver. In particular, this analysis assumes a Gaussian distribution of the output of the PDs, and incorporates the combined effects of atmospheric turbulence, pointing error induced geometrical loss, the size of the PDs in the array, and AoA fluctuations of the received optical beam. We provide simulation results to verify the accuracy of the derived analytical expressions and study the effect of hovering fluctuations on the system performance. Indeed, simulation results show that the performance of such links depends heavily  on the random fluctuations of the hovering UAV. Also, it is shown that enlarging the receiver field-of-view via employing the array of detectors can help alleviate the adverse effects of random hovering fluctuations on the link performance. Furthermore, the results show that for each level of instability there is an optimal size for the array that  minimizes the tracking error probability. Since estimating the beam position at the receiver is part of the channel information required for data detection, minimizing tracking error will ultimately improve system performance in terms of BER.
 }

\textcolor{black}{To the best of
authors’ knowledge, there is little work in the literature that
mathematically models beam tracking error under random orientation
fluctuations by taking into account the effects of all channel impairments in an intensity modulation direct detection (IM/DD) FSO system.}
\section{System Model}
\label{II}
{

\subsection{Signal Model}

We consider an IM/DD FSO system with on-off keying signaling. The received signal at the $i$th quadrant corresponding to the $k$th symbol interval is given as 
\begin{equation}
\label{photocurrent}
r_{i}[k] =  h D_i s[k] + n_i[k], \text{~~~for~~~} i \in \{1, \ldots, 4\}
\end{equation}
where $s[k]$  denotes the transmitted symbol with optical power $P_t$, and $h$ denotes the fading channel coefficient, which is assumed to be constant over a large sequence number of transmitted bits (i.e., quasi static property). Let us define $D_i \in \{0,1\}$ as a signal indicator, i.e, $D_i = 1$ indicates that the received beam is captured by the $i$th quadrant. It is worth noting that, in practice, the size of PDs is on the order of several millimeters, being much larger than the airy width of the received optical beam. This beam is approximately equal to $2.4\lambda$, where $\lambda$ is the optical wavelength and lies in the range of a few hundred nanometers. Hence, it is reasonable to ignore the effect of boundary conditions on the four-quadrant detector \cite{gagliardi1995optical,kiasaleh2006beam}. As a result, if the deviated received beam is still within the receiver FoV, it will be captured by the $i$th quadrant of the PD. Moreover, $n_i[k]$ in \eqref{photocurrent} denotes the noise of the $i$th quadrant and it is an additive white Gaussian noise (AWGN) having mean zero and variance 
\begin{align}
\label{variance-noise-i,k}
\sigma_{i,k}^2 = \sigma_s^2 h D_i s[k] + \sigma_0^2, \text{~~~for~~~} i \in \{1, \ldots, 4\}
\end{align}
where $\sigma_s^2$ denotes the variance of the shot noise due to transmitted signal; We have $\sigma_0^2 = \sigma_{b}^{2} + \sigma_{th}^{2}$ where $\sigma_{b}^{2}$ and $\sigma_{th}^{2}$ denote the noise variance due to undesired background radiation and receiver thermal noise, respectively.  Let us define $\Omega_{FoV}$ as the solid angle FoV of the receiver; therefore, the collected background power, $P_b$, is obtained as \cite{j2018channel}
	\begin{align}
	\label{xf1}
	P_b = N_b(\lambda)\, B_o\, \Omega_{FoV}\, A_a
	\end{align}
	where $N_b(\lambda)$ (in Watts/${\rm cm}^2$-$\micro$m-srad) denotes the spectral radiance of the background radiations at wavelength $\lambda$ , $B_o$ (in $\micro$m) denotes the bandwidth of the optical filter at the Rx , and $A_a$ (in ${\rm cm}^2$) denotes the lens area. To calculate $\Omega_{FoV}$, we denote  $\theta_{FoV} = \frac{2r_a}{f_c}$ by the range of optical beam arrival angles observed by the PD area. \textcolor{black}{Here, $r_a$  denotes the radius of the quadrants, and $f_c$ denotes the focal length of the receiver lens.} Accordingly, $\Omega_{FoV}$ can be obtained as 
\small
	\begin{align}
	\label{circle_fov}
	\Omega_{FoV} = \int_0^{\frac{\pi}{2}}        \int_0^{\frac{\theta_{FoV}}{2}} \sin(\theta)d\theta d\phi 	= \dfrac{\pi}{2}\left(1-\cos\left(\frac{\theta_{FoV}}{2}\right)\right) \simeq \frac{\pi{r_a^2}}{4f_c^2}.
	\end{align}
\normalsize
Hence, in the considered setup, $P_b$ is obtained as
	\begin{align}
	\label{Pbb}
	P_b =   \frac{\pi r_a^2\, N_b(\lambda)\, B_o\,A_a}{4f_c^2}. 
	\end{align}
\subsection{FSO Channel Model}
The channel coefficient can be modeled as $h = h_lh_ah_p$, where $h_l$ is the deterministic attenuation loss, $h_a$ denotes the atmospheric turbulence, and $h_p$ is the pointing error. 
The probability density function (PDF) of $ h $ is given by \cite{sandalidis2009optical}

\small
\begin{align}
\label{fading}
f_{h}(h)=  \frac{\alpha\beta\gamma^{2}}{A_{0}h_{l}\Gamma(\alpha)\Gamma(\beta)}         
	\times G_{1,3}^{3,0}   \left( \frac{\alpha\beta}{A_{0}h_{l}}h  \  \Bigg\vert \  {\gamma^{2} \atop \gamma^{2}-1,\alpha-1,\beta-1} \right)
\end{align}
\normalsize
where $  G_{1,3}^{3,0}   \left( \cdot \right) $ and $\Gamma(\cdot)$ denote the Meijer\textquotesingle  s $  G$ function and the Gamma function, respectively. The parameter $ \gamma$ denotes the ratio between the equivalent beam radius at the receiver and the pointing errors jitter, and the parameter $A_0$ is the maximal fraction of the collected power.
Furthermore, $  1/\beta$ and $1/\alpha$ are, respectively, the variances of the small scale and large scale eddies, and they can be determined using the Rytov variance \cite{sandalidis2009optical}. 
\subsection{Beam Deviation Due to The Hovering Fluctuations}
\begin{figure}
	\begin{center}
		\includegraphics[width=5 in]{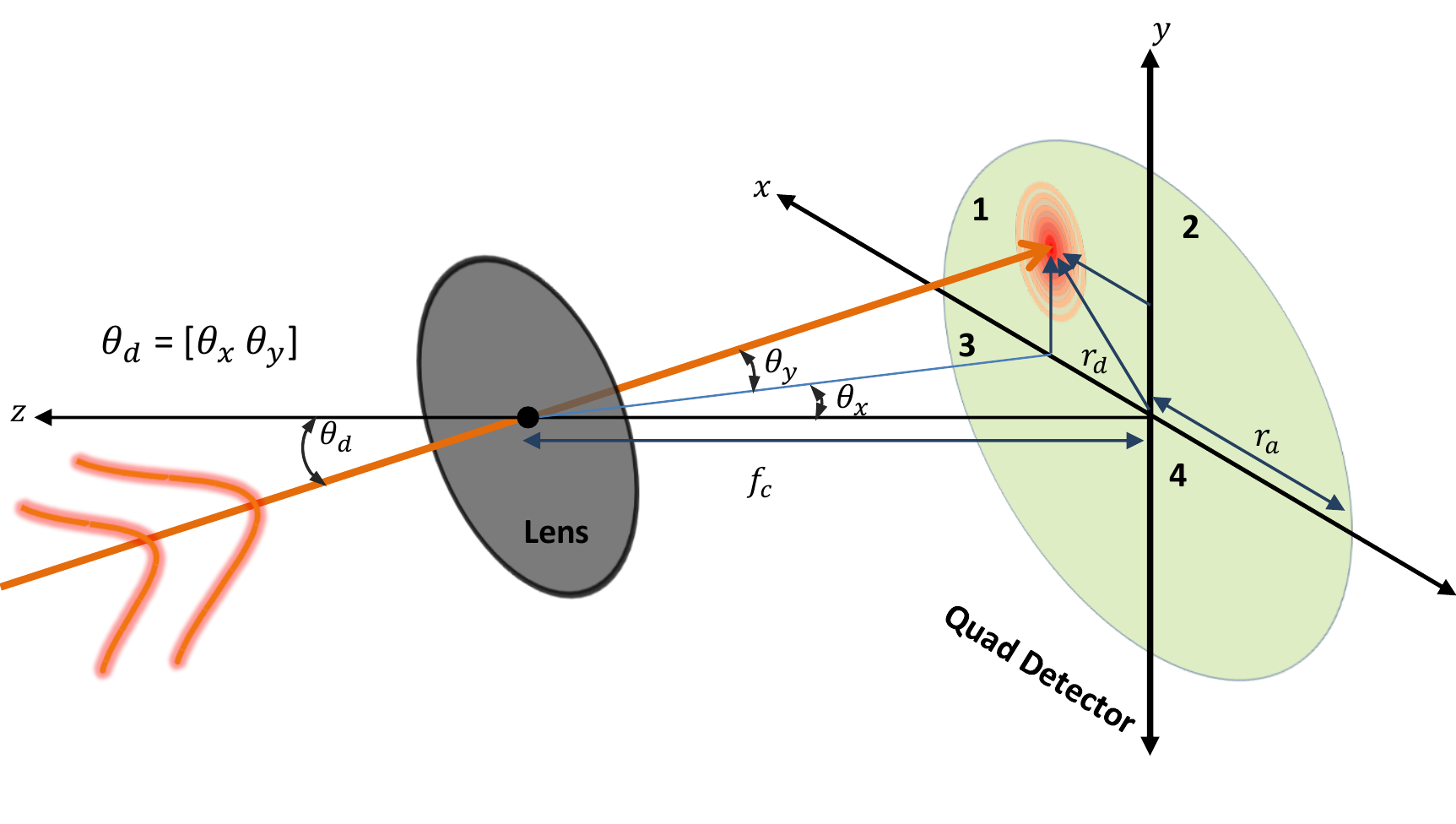}
		\caption{The schematic of deviated received beam due to the UAV's wobbling at the plane of the four-quadrant detector is depicted. \textcolor{black}{Here, $f_c$ is the focal length of the receiver lens, $r_a$ is the radius of the quadrants, and $r_d$ is the radial beam displacement at the detector array. Furthermore, $\theta_x$ and $\theta_y$ represent the orientation deviations of received optical beam at $x-z$ and $y-z$ planes, respectively.   }}
		\label{b}
	\end{center}
\end{figure}

As depicted in Fig. \ref{b}, we define two independent random variables (RVs), namely $\theta_x$ and $\theta_y$, to model the orientation deviations of received optical beam at the receiver plane. According to \cite{j2018channel}, these two RVs are Gaussian distributed with mean zero and variances  $\sigma_x^2$ and $\sigma_y^2$, when  their joint PDF is obtained as
\begin{align}
\label{orientation}
p_{\theta}\left( \theta_x,\theta_y\right) = \frac{1}{2\pi \sigma_x\sigma_y} \exp\left( -\frac{\theta_x^2}{\sigma_x^2}  -\frac{\theta_y^2}{\sigma_y^2}\right).
\end{align}
The full beam misalignment will occur  if the orientation of the received optical beam, $\theta_d = \sqrt{\theta_x^2+\theta_y^2}$, deviates bigger than the receiver FoV. Such condition implies $D_i=0$ for $i\in\{1, ..., 4\}$. The occurrence probability of full beam misalignment is defined  as
\begin{align}
	\label{interrupt}
	{P}_{f}= 1 - \sum_{i=1}^4 {P}_{D_i}
\end{align}
where ${P}_{D_i}$ is the probability of capturing the arrival beam at the $i$th quadrant. Due to the symmetric shape of the detector we have  
${P}_{D_i}= {P}_{D_j}$ for $i$ and $j\in\{1,...,4\}$, and, for example, ${P}_{D_1}$ can be attained as
{\small
\begin{align}
\label{pd1d}
{P}_{D_1} &= \int_0^{\tan^{-1}\left(\frac{r_a}{f_c}\right)} \int_0^{\tan^{-1}\left(\frac{r_a}{f_c}\right)}  p_{\theta}\left( \theta_x,\theta_y\right)  d\theta_xd\theta_y  \\
&=\left(\frac{1}{2}-Q\left( \frac{\tan^{-1}\left(\frac{r_a}{f_c}\right)}{\sigma_x}\right) \right)
  \left(\frac{1}{2}-Q\left( \frac{\tan^{-1}\left(\frac{r_a}{f_c}\right)}{\sigma_y}\right) \right). \nonumber
\end{align}
}
\normalsize
\section{Beam Tracking}
\label{III}
The received data is gathered during an observation window comprised of $L_s$ bits, i.e.,  $\underline{\mathbf{r_i}} = \{r_{i}[1],r_{i}[2],\ldots,r_{i}[L_s]\}$ at the $i$th quadrant of the quad-detector and corresponding to the transmitted signal vector $\underline{\mathbf{s}} = \{s[1],s[2],\ldots,s[{L_s}]\}$. Due to the slow fading characteristic of the FSO channel, the channel fading $h$ can be estimated at the receiver using either pilot symbols or blind methods.

Let $m = \sum\limits_{k = 1}^{{L_s}} {s[k]}$ be the number of bits `1' in the vector of transmitted signal. Therefore, at the $i$th quadrant, the received signal conditioned on $h$ and $m$ can be written as 
\begin{align}
\label{conditioned-r}
{r'_{i|h,m}} = \sum\limits_{k = 1}^{{L_s}} {{r_i}[k] = h{D_i}}  m + {n'_{i|h,m,{D_i}}}
\end{align}
where ${n'_{i|h,m,{D_i}}} = \sum\limits_{k = 1}^{{L_s}} {n_{i}[k]}$ denotes the AWGN with mean zero and variance as
\begin{align}
\label{variance-noise-i}
\sigma_{i|h,m,D_i}^2 = \sigma_s^2 h D_i m + L_s\sigma_0^2.
\end{align}                     
Thus, the PDF of ${r'_{i|h,m}}$ conditioned on $D_i$ can be written as
{\small
\begin{align}
\label{pdf-r-conditional}
p(r'_{i|h,m}|{D_i}) \!=\! \frac{1}{{\sqrt {2\pi \sigma _{i|h,m,{D_i}}^2} }}\exp \!\!\left( - {\frac{{{{\left( {r{'_{i|h,m}} - h{D_i} m} \right)}^2}}}{{2\sigma _{i|h,m,{D_i}}^2}}} \right)\!\!.
\end{align}}
For optimum beam tracking, the decision on selecting the  $i$th quadrant as the target quadrant capturing the received optical beam is made based on the maximum likelihood (ML) criterion as follows
\small
\begin{align}
\label{ML-criterion}
\hat i =& \underset{i \in \{1,\ldots,4\}}{\operatorname{\text{~arg~max~}}}~p(r'_{i|h,m}|{D_i=1})\times \!\!\!\!\!\!\prod_{j=1,j\ne i}^4 \!\!\!\!p(r'_{j|h,m}|{D_j=0}) \nonumber \\ 
=& \underset{i \in \{1,\ldots,4\}}{\operatorname{\text{~arg~max~}}} \log\left(p(r'_{i|h,m}|{D_i=1})\right)+ \!\!\!\!\!\sum_{j=1,j\ne i}^4 \!\!\!\!\log\left(p(r'_{j|h,m}|{D_j=0})\right). 
\end{align}
\normalsize
Substituting (\ref{pdf-r-conditional}) into (\ref{ML-criterion}), after some mathematical derivations, we can express the optical beam tracking based on the metric $\mathcal{T}_{i|h,m}$ as
\begin{align}
\label{metric-tracking-1}
\hat i = \underset{i \in \{1,\ldots,4\}}{\operatorname{\text{~arg~min~}}}~\mathcal{T}_{i|h,m}
\end{align}
where
\begin{align}
\label{metric-tracking-2}
\mathcal{T}_{i|h,m} = \frac{{{{\left| {{{r'}_{i|h,m}} - h m} \right|}^2}}}{{\sigma _s^2hm + {L_s}\sigma _0^2}} +\!\!\! \sum_{j=1,j\ne i}^4 \!\!{\frac{{{{\left| {{{r'}_{j|h,m}}} \right|}^2}}}{{{L_s}\sigma _0^2}}}. 
\end{align}
The probability of tracking error for the proposed method is derived in Appendix A as 
\small
\begin{align}
\label{TER_PCSI}
&{P}^{\rm p}_{te} \simeq {P}_{f} +  \frac{(1-{P}_{f})}{2^{L_s}}  \int_0^\infty \sum_{m=0}^{L_s} \binom{L_s}{m} \nonumber \\ &\Bigg\{1- 
\left( 1-Q\left(  \frac{ \sigma_s^2 h^2m^2 \left(h m+2L_s\sigma_0^2\right)}
{\sigma_{tc|h,m}}  \right) \right)^3 \Bigg\}f_h(h)dh
\end{align}
\normalsize
where
\begin{align}
\label{hjj3}
\sigma_{tc|h,m}^2  =&~\left(2\sigma_s^2 h m \left(h m + L_s\sigma_0^2 \right)\right)^2 
\cdot \left(\sigma_s^2 h m + L_s\sigma_0^2\right)  \nonumber \\ 
&~+L_s\sigma_0^2   \left(2mL_s\sigma_0^2 \sigma_s^2 h\right)^2  .
\end{align}

\section{Simulation Results and Discussion}
\label{V}
\begin{figure}[t]
	\begin{center}
		\includegraphics[width=5in]{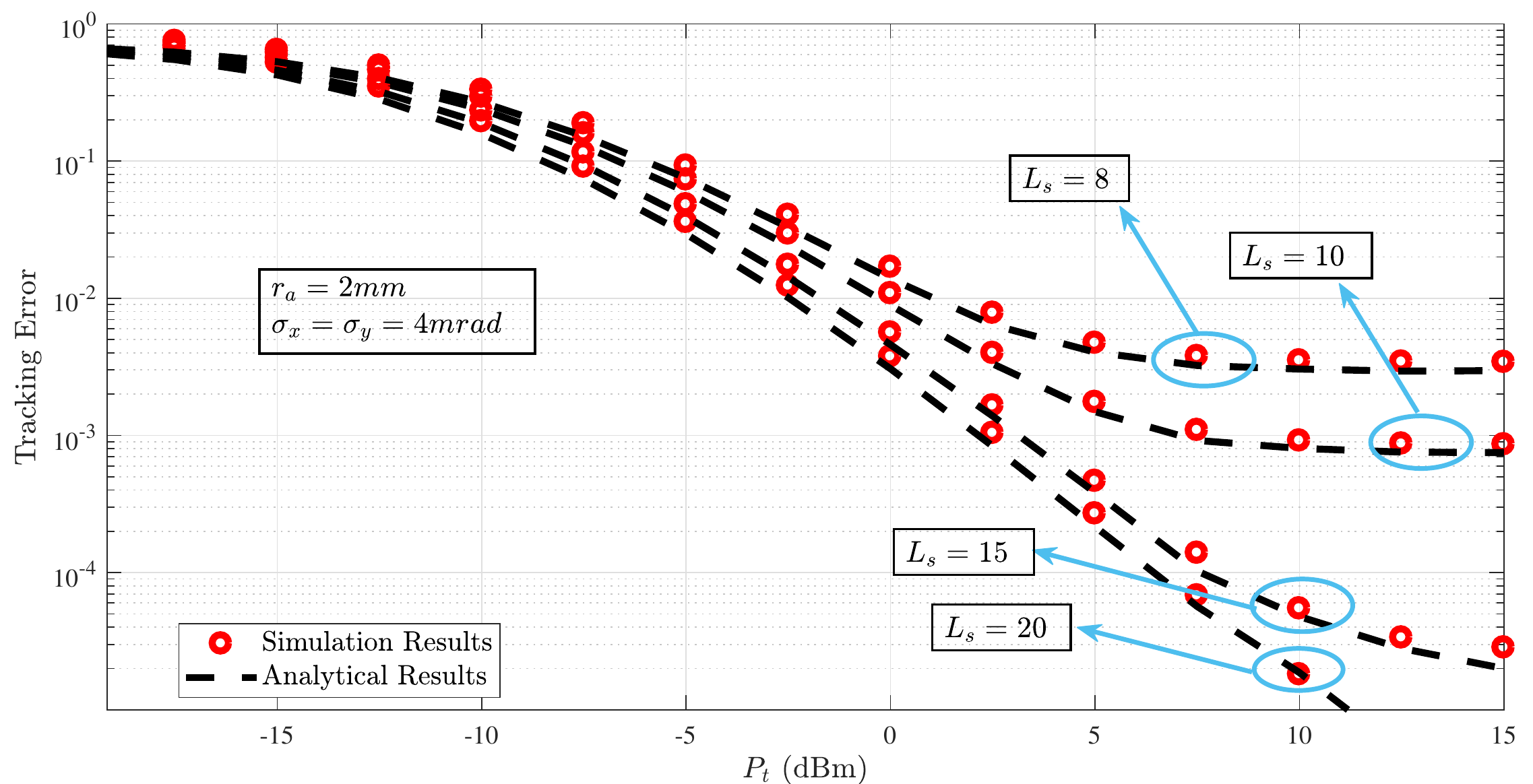}
		\caption{Tracking error versus $P_t$ for different values of $L_s$}
		\label{tracking_different_Ls}
	\end{center}
\end{figure}
\begin{figure}[t]
	\begin{center}
		\includegraphics[width=5 in]{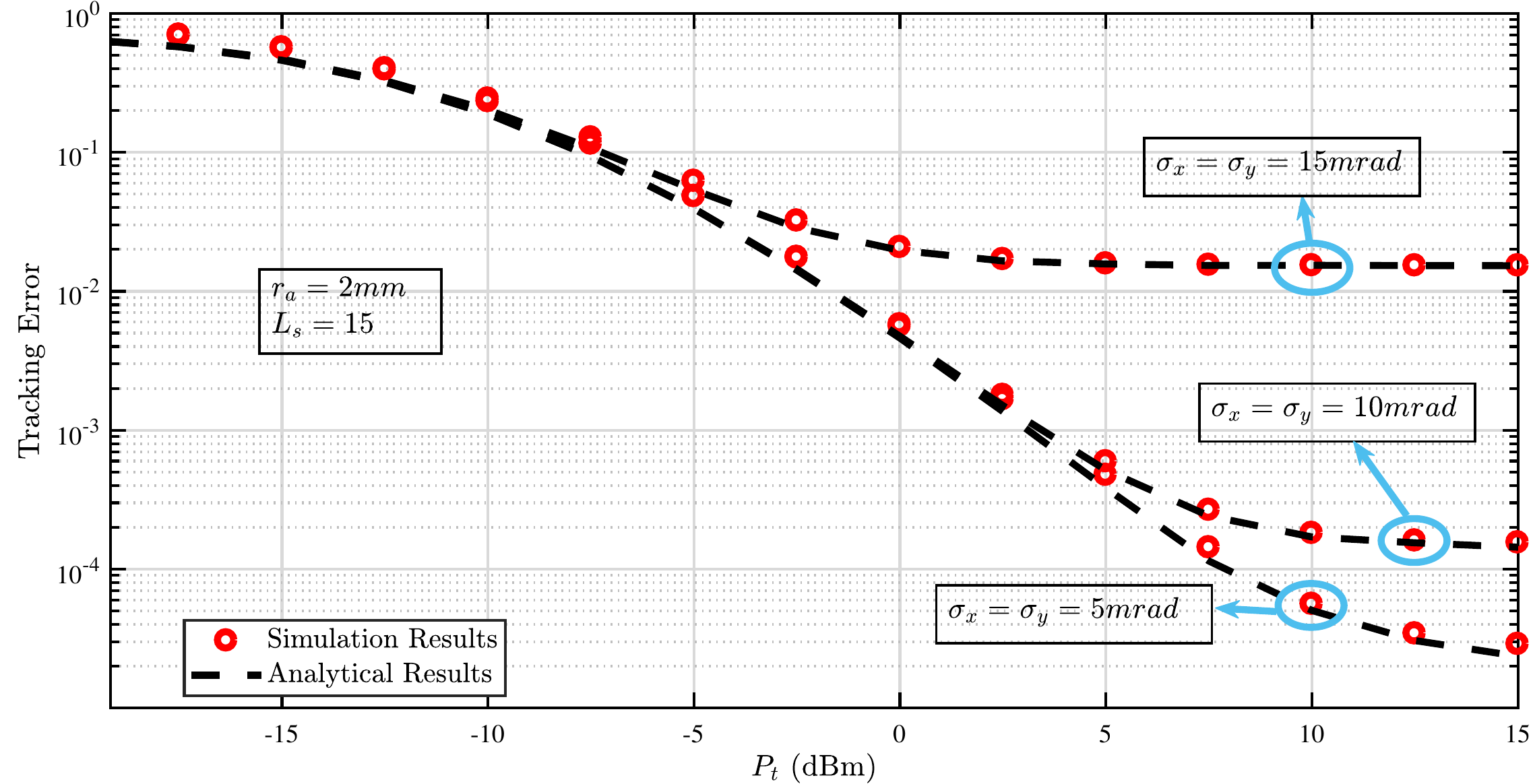}
		\caption{Tracking error versus $P_t$ for different values of $\sigma_x$ and $\sigma_y$}
		\label{tracking_different_sigma}
	\end{center}
\end{figure}

\begin{figure}[t]
	\begin{center}
		\includegraphics[width=5 in]{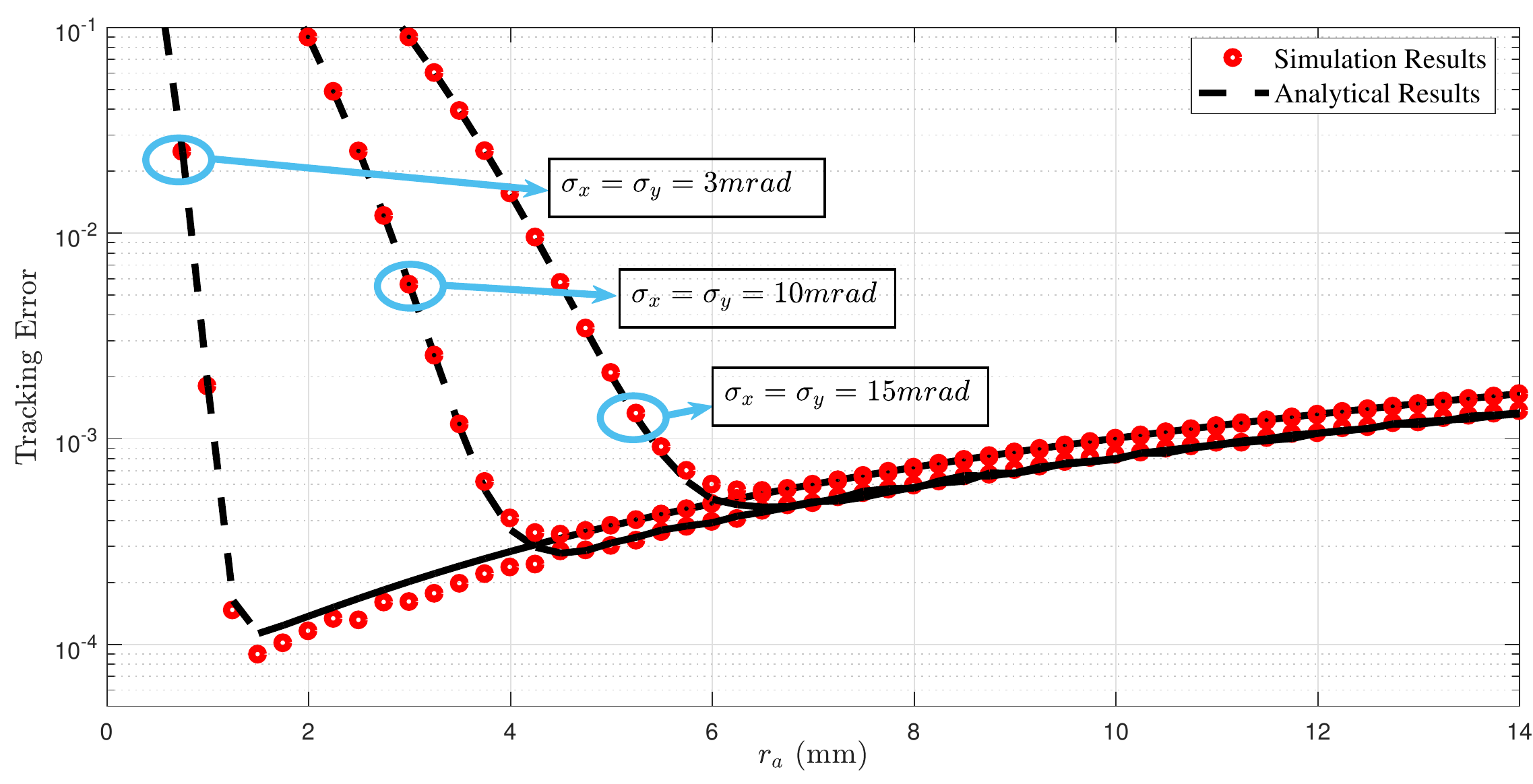}
		\caption{Tracking error versus detector size for different values of $\sigma_x$ and $\sigma_y$}
		\label{tracking_vs_size}
	\end{center}
\end{figure}
\begin{figure}[t]
	\begin{center}
		\includegraphics[width=5 in]{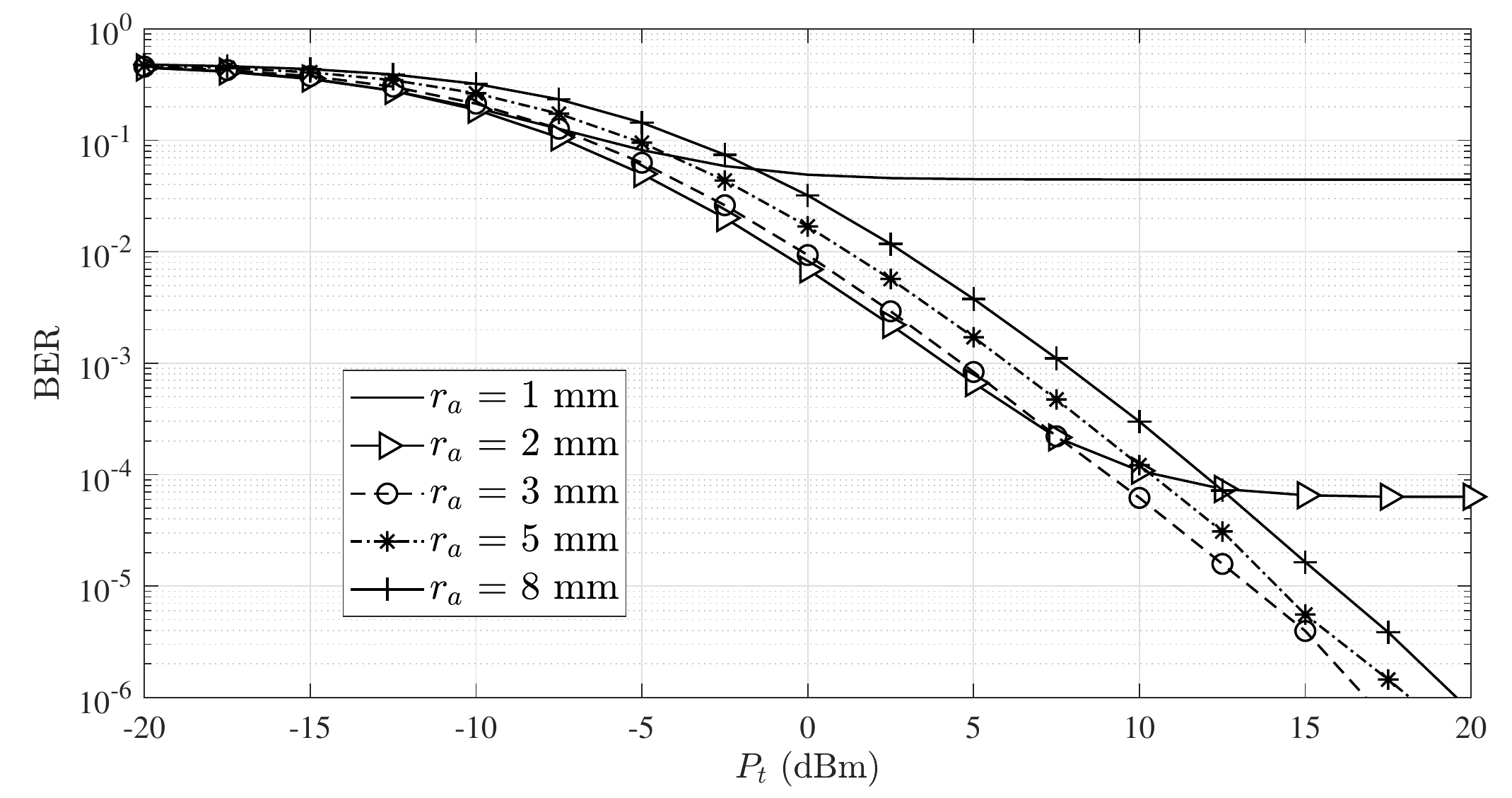}
		\caption{\textcolor{black}{BER versus $P_t$ for different values of detector size $r_a$, when $\sigma_x$ = $\sigma_y$ = $5 \textrm{~mrad}$, $L_s = 20$} }
		\label{BER_vs_power}
	\end{center}
\end{figure}

We provide simulation results to investigate the effect of hovering fluctuations and the detector size on the performance of considered system. Moreover, the  accuracy of analytical analysis is corroborated by performing $6\times10^6$ independent runs through Monte-Carlo simulations.   \textcolor{black}{For simulations, we set the parameter values as follows \cite{j2018channel}. Rytov variance $= 1$, $\text{aperture radius} = 5$ cm, $f_c = 5$ cm, $N_b(\lambda)\ = 10^{-3}$ W/$\text{cm}^2$-m-srad, $B_o = 10$ nm, and $A_0 = 0.0198$.} Fig. \ref{tracking_different_Ls} depicts tracking error versus $P_t$ for different values of $L_s$. As we can observe, increasing $L_s$ improves the performance of the tracking method at the expense of more delay of tracking. However, an error floor for tracking error is realized due to the transmission of all-zero sequence having the occurrence probability ${1}/{2^{L_s}}$. Nevertheless, this error can be avoided by adopting a source preceding method at the transmitter \cite{7433927}.   Furthermore, the dramatic impact of hovering fluctuations on the proposed tracking approach is reflected in Fig. \ref{tracking_different_sigma}. 
However, this negative impact can be compensated by enlarging the size of the PD at the receiver, which increases the Rx FoV. 

\textcolor{teal}{Moreover, Fig. \ref{tracking_vs_size} plots tracking error versus detector size for different values of hovering fluctuations. As shown, widening the receiver FoV by using larger PDs can help compensate tracking error  at the expense of less electrical bandwidth and accept more undesired background noise at the receiver. Meanwhile, Fig.   \ref{tracking_vs_size} demonstrates that for each level of instability there exist an optimal size of the detector that can minimizes the tracking error probability.}  \textcolor{black}{Furthermore, to demonstrate the impact of  detector size on
	the overall system performance, we plot
	the BER versus transmit power for different radius of the quadrants in Fig. {\ref{BER_vs_power}}. Again, it can be seen from this figure that increasing
	the size of the quadrants can help mitigate the adverse effect of AoA fluctuations on the system performance. However, from a certain size onward ($r_a = 5 \textrm{~mm}$ in this figure), further increase does not necessarily improve the
	system performance. This observation is expected, because increasing the detector size
	results in accepting more undesired background noise due to a wider receiver FoV.  Furthermore, for the small sizes of the detector and from a certain level of transmit power onward, increasing the power does not necessarily have a noticeable effect on improving the system performance. In this case, the system performance is limited to the tracking error, and when the detector size is small, the amount of this error cannot be reduced to an acceptable level, resulting in an error floor.}

\section{Conclusion}
We investigated the effect of random hovering fluctuations on the performance of beam tracking method for an FSO link to a UAV.  Extensive mathematical analysis was carried out, and a semi-closed form expression for the tracking error probability was derived. Our analytical method can find the optimal detector size for minimizing tracking error without resorting
to time-consuming simulations.
\appendices
\section{Tracking Error Analysis}
For the proposed method, tracking error is expressed as 
\begin{align}
\label{kd111}
P_{te} = {P}_{f} + (1-{P}_{f})\int_0^\infty {P}_{te|h} f_h(h) dh
\end{align}
where 
\begin{align}
\label{kd222}
{P}_{te|h} = \sum_{m=0}^{L_s} p(m) {P}^{\rm p}_{te|h,m}
\end{align}
\begin{align}
\label{kd3}
{P}^{\rm p}_{te|h,m} = 1-{P}_{tc|h,m}^{\rm p}.
\end{align}
We have ${P}^{\rm p}_{te|h,m}$ and ${P}^{\rm p}_{tc|h,m}$ as the tracking error probability  and the probability of correct tracking conditioned on $h$ and $m$, respectively. Furthermore, $ P(m) = \binom{L_{s}}{m}/2^{L_s} $ denotes the probability that $ m $ bits out of $ L_s $ transmitted bits are equal to one where $\binom{n}{m}$ is the number of combinations of $m$ items out of $n$ items. Without loss of generality, we assume that the first quadrant is the target PD, i.e., $D_1 = 1$. Then, we have
\begin{align}
\label{kd4}
{P}_{tc|h,m}^{\rm p} = {\rm Prob}\left\{\mathcal{T}_{1|h,m} < \mathcal{T}_{\min|h,m} \right\}
\end{align}
where 
\begin{align}
\label{kdd1}
\mathcal{T}_{\min|h,m} = \min\left(\mathcal{T}_{2|h,m},\mathcal{T}_{3|h,m},\mathcal{T}_{4|h,m}\right).
\end{align}
The noise of the PDs are independent; therefore \eqref{kd4} simplifies to
\begin{align}
\label{kd5}
{P}_{tc|h,m}^{\rm p} = \left({P}'^{\rm p}_{tc|h,m}\right)^3
\end{align}
where
\begin{align}
\label{kd6}
{P}'^{\rm p}_{tc|h,m} &= {\rm Prob}\left\{\mathcal{T}_{1|h,m} < \mathcal{T}_{i|h,m} \right\}  \text{~for~} i \in {2,3,4}.
\end{align}
Substituting \eqref{metric-tracking-2} into \eqref{kd6}, we can express ${P}'^{\rm p}_{tc|h,m}$ as \eqref{kd7} at the bottom of this page.
%
\small
\begin{figure*}[b]
	\normalsize
	\hrulefill
	\begin{align}
	\label{kd7}
	{P}'^{\rm p}_{tc|h,m} &= {\rm Prob}\left\{ 
	\frac{{{{\left| {{{r'}_{1|h,m}} - h m} \right|}^2}}}{{\sigma _s^2hm + {L_s}\sigma _0^2}}
	+\sum_{j=2}^4\!{\frac{{{{\left| {{{r'}_{j|h,m}}} \right|}^2}}}{{{L_s}\sigma _0^2}}} <
	\frac{{{{\left| {{{r'}_{2|h,m}} - h m} \right|}^2}}}{{\sigma _s^2hm + {L_s}\sigma _0^2}} +
	\frac{{{{\left| {{{r'}_{1|h,m}}} \right|}^2}}}{{{L_s}\sigma _0^2}}
	+\sum_{j=3,4}{\frac{{{{\left| {{{r'}_{j|h,m}}} \right|}^2}}}{{{L_s}\sigma _0^2}}} 
	\right\} \nonumber \\
	&= {\rm Prob}\left\{ 
	\sigma_s^2hm \left(  \left( r'_{1|h,m}  \right)^2  -  \left( r'_{2|h,m}  \right)^2 \right)
	+ 2mL_s\sigma_0^2 \sigma_s^2 h \left(   r'_{1|h,m}  -  r'_{2|h,m}  \right) >0
	\right\}
	\end{align} 
\end{figure*}
\normalsize
From \eqref{conditioned-r},  eq. \eqref{kd7} is rewritten as
\small
\begin{align}
\label{kd12}
&{P}'^{\rm p}_{tc|h,m}  
=\!{\rm Prob}\bigg\{\!  
\sigma_s^2hm \!\left(  \left( h m + n'_{1|h,m,D_1=1}\right)^2 \right. 
\!\!-\! \! \left. \left( n'_{2|h,m,D_2=0}  \right)^2 \right) \nonumber \\
&~~+ \!2mL_s\sigma_0^2 \sigma_s^2 h \left( h m + n'_{1|h,m,D_1=1}- n'_{2|h,m,D_2=0} \right) >0
\bigg\} \nonumber \\
&=\!{\rm Prob}\bigg\{\!
 \sigma_s^2 h^2m^2\!\left(h m+2L_s\sigma_0^2\right)\! + \!n'_{tc|h,m} >0\bigg\}
\end{align}
\normalsize
where
\small
\begin{align}
\label{hj1}
n'_{tc|h,m} =&~ \sigma_s^2hm \left(  \left( n'_{1|h,m,D_1=1}\right)^2  
- \left( n'_{2|h,m,D_2=0}  \right)^2 \right) \nonumber \\
&+2\sigma_s^2 h m \left(h m + L_s\sigma_0^2 \right) n'_{1|h,m,D_1=1} \nonumber \\
&-2m L_s\sigma_0^2\sigma_s^2 h  n'_{2|h,m,D_2=0}.
\end{align}    
\normalsize              
At high SNR, we have $\left( n'_{1|h,m,D_1=1}\right)^2\ll n'_{1|h,m,D_1=1}$ and 
$\left( n'_{2|h,m,D_2=0}\right)^2\ll n'_{2|h,m,D_2=0}$. Hence,  eq. \eqref{hj1} can be approximated as
\begin{align}
\label{hj2}
n'_{tc|h,m} \simeq&~ 2\sigma_s^2 h m \left(h m + L_s\sigma_0^2 \right) n'_{1|h,m,D_1=1}  \\
&-2m L_s\sigma_0^2\sigma_s^2 h  n'_{2|h,m,D_2=0}.\nonumber
\end{align} 					
From \eqref{hj2} and \eqref{variance-noise-i}, $n'_{tc|h,m}$ can be described by a Gaussian distribution with mean zero and variance 						
\begin{align}
\label{hj3}
\sigma_{tc|h,m}^2  =&~\left(2\sigma_s^2 h m \left(h m + L_s\sigma_0^2 \right)\right)^2 
\times \left(\sigma_s^2 h m + L_s\sigma_0^2\right)  \nonumber \\ 
&~+L_s\sigma_0^2   \left(2mL_s\sigma_0^2 \sigma_s^2 h\right)^2  .
\end{align}
Based on \eqref{hj3} and \eqref{kd12}, ${P}'^{\rm p}_{tc|h,m}$ is derived as
\begin{align}
\label{hj4}
{P}'^{\rm p}_{tc|h,m} \simeq 1-Q\left(  \frac{ \sigma_s^2 h^2m^2 \left(h m+2L_s\sigma_0^2\right)}
{\sigma_{tc|h,m}}  \right).
\end{align}
Finally, by substituting \eqref{hj4}, \eqref{kd5}, \eqref{kd3}, and \eqref{kd222} into \eqref{kd111}, we obtain the analytical  expression of  tracking error probability in \eqref{TER_PCSI}.
\bibliographystyle{IEEEtran}


\end{document}